\documentclass[10pt,aps,prb,superscriptaddress,raggedbottom,showpacs,floatfix, twocolumn]{revtex4-1}

\usepackage{amsmath,amsfonts,amssymb,amsthm}
\usepackage{graphicx,latexsym}
\usepackage{bm}

\begin{document}

\title{Effects of scattering area shape on spin conductance in a four-terminal spin-Hall setup}
\author{Gunnar Thorgilsson}
\affiliation{Science Institute, University of Iceland, Dunhaga 3, IS-107 Reykjavik, Iceland}
\author{Sigurdur I. Erlingsson}
\affiliation{Reykjavik University, School of Science and Engineering, Menntavegi 1, IS-101 Reykjavik, Iceland}

\begin{abstract}
We study spin conductance in a ballistic and quasi-ballistic two dimensional electron system
with Rasbha spin-orbit coupling. The system has a four-terminal geometry with round corners at
the connection to the leads. It is found that by going from sharp corners to more round
corners in the ballistic system the energy depended spin
conductance goes from being relatively flat to a curve
showing a series of minima and maxima.
It is also found that when changing the size of the terminal area by modifying the
roundness of the terminal corners the maxima and minima in the transverse spin conductance are
shifted in energy. This shift is due increased (decreased) energy for smaller (larger) terminal
area. These results were also found to be reasonably stable in quasi-ballistic systems.
\end{abstract}


\maketitle

\section{Introduction and motivation}
Spin-orbit (SO) coupling in semiconductor nanostructures has for some time been considered as one of the 
main candidates in controlling spin in semiconductor spintronic devices
\cite{Wolf:science2001,Awschalom:Book2002,Awschalom:physnature2007}. One of the main benefits of the 
SO coupling is that it can be used to manipulate electron spins in semiconductors through gate voltages
instead of external magnetic fields\cite{Prinz:science1998} or magnetic doping\cite{Dietl:book1994}.
SO coupling arises from relativistic effects and when combined with $\bm{k}\cdot \bm{p}$\
-\ calculation it will lead to spin-orbit terms depending on the crystal structure, i.e.\
 the Dresselhaus SO coupling\cite{Dresselhaus:pr1955,Winkler:book2010} in crystals that lack inversion symmetry.
In addition, further spin-orbit contributions can occur at heterostructure interfaces.
This contribution is the so-called Rashba SO coupling\cite{Rashba:spss1960,Winkler:book2010}. 
Voltages applied to local gates can change the heterostructure confining potential, thus modifying
the Rashba SO coupling\cite{Nitta:prl1997,Schapers:sst2009}. More recently, yet a new type of spin-orbit
interaction has been found in symmetric two-dimensional quantum structures with two subbands: the inter-subband
induced spin-orbit coupling\cite{Bernardes:prl2007,Calsaverini:prb2008}. 

In the field of spintronics the main goals is the creation and detection of spin-currents.  In 
electron doped semiconductor structures one of the main candidates for spin-current creation is the so-called
spin-Hall effect\cite{Sinova:prl2004}.  Although it has been proved that in extended 2D systems the spin-Hall effect
vanishes \cite{Inoue:prb2004,Chalaev:prb2005,Erlingsson:prb2005}, this results does not hold in finite size systems.
The exception here are systems with the inter-subband induced spin-orbit coupling\cite{Bernardes:prl2007,Calsaverini:prb2008},
which gives rise to a non-zero spin Hall effect\cite{Lee:prb2009}, even for extended systems.
The interplay of in-plane confinement and spin-orbit interaction leads to many interesting spin-related transport phenomena.

The Rashba SO coupling has been in recent years extensively studied both experimentally
\cite{Miller:prl2003,Schapers:prb2004,Kato:science2004,Sih:naturephysics2005,Sih:prl2006,Schapers:sst2009,Debray:naturenano2009}
and theoretically, either in two-terminal setups,
\cite{Reynoso:prb2006,Zhang:prb2005,Nicolic:prb2005:a,Nikolic:prl2005,Nikolic:prb2005:b,Duckheim:prb2009},
or in multiterminal setups
\cite{Garelli:prb2009,Xing:prb2006,Yokoyama:prb2006,Yamamoto:prb2005,Nikolic:prl2005,Nikolic:prb2005:c}.
Previous numerical studies hint that the shape of the scattering area of the system plays a 
vital role in the behavior of the spin current flowing through the transverse leads 
\cite{Xing:prb2006,Yokoyama:prb2006}. Also one expects that gate defined nanostructures will have confinement potential that changes on a length
scale much larger than the Fermi wavelength of the electrons\cite{Ihn:book2010}.

In this work we will present calculation of spin conductance through transverse leads in a four
terminal spin Hall setup where the shape of the scattering region is changed. Most lattice model
calculations\cite{Garelli:prb2009,Xing:prb2006,Yokoyama:prb2006,Yamamoto:prb2005,Nikolic:prl2005,Nikolic:prb2005:c}
used abrupt edges  in the way the leads where connected to the scattering region. This abruptness leads
to substantial scattering (both of charge and spin) which can suppress spin-related phenomena that one is interested in.
We use smooth connection of leads to the scattering area. Since most semiconductor heterostructures
are defined by gates, one would expect relatively smooth confining potentials \cite{Ihn:book2010}.
We study the spin conductance as a function of scattering area shape and propose an explanation of
the observed connection between the spin conductance and the change of size of the scattering region.
We also observe that the values of spin conductance is relatively large which is due to the smooth
transfer of electrons from the longitudinal leads to the transverse one. 

The paper is organized as follows. In section II we present the theory behind the calculations. Section III
defines the parameters used in the calculations and presents the results for the ballistic system in
subsection A and for the quasi-ballistic system in subsection B. Finally section IV contains
conclusions and discussion.

\section{Theory}
\begin{figure}[htp]
 \begin{center}
 \includegraphics[width=0.45\textwidth]{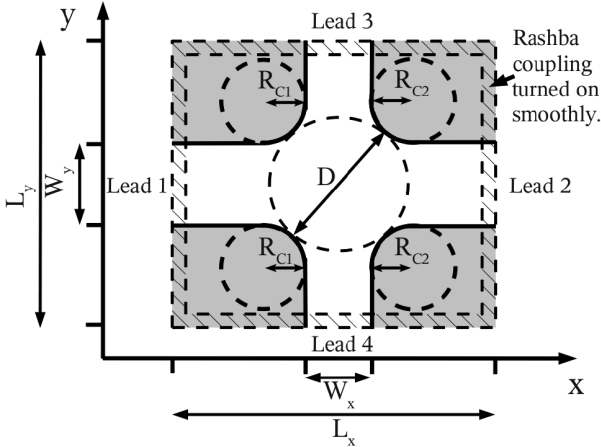}
 \end{center}
 \caption{
 Schematic picture of the four terminal system. Sharp corners are replaced
 by round corners with radii $R_{\mathrm{C}1}$ and $R_{\mathrm{C}2}$, see Fig.\ \ref{fig:scematic}. The 
 whole system has a length of $L_x=102 a=600$ nm and width of $L_y=L_x$.
 Leads 1 and 2 are the main leads into the system with a width of
 $W_y=41 a = 241$ nm, while leads 3 and 4 are the transverse leads
 into the system with a width of $W_x= 18 a = 106$ nm. The Rashba
 SO coupling is turned on smoothly on a $10 a = 58.8$ nm thick
 strip framing the system (hatched area in the figure). Inside the scattering region
 is shown a circle of width $D$ which we use to estimate the size of the terminal area.
 } 
 \label{fig:scematic}
\end{figure}

We are interested in a four terminal 2DEG in a semiconductor heterostructure. A schematic of the
scattering region of the system can be seen in Fig.\ \ref{fig:scematic}. The system is described by
the following effective mass Hamiltonian
\begin{equation}
 H(x,y)=\frac{p_x^2+p_y^2}{2m^*}+V_{\mathrm{C}}(x,y)+H_{\mathrm{R}}(x,y),
 \label{eq:MainHamilton}
\end{equation}
where $m^*$ is the effective electron mass.
$V_{\mathrm{C}}(x,y)$ describes a hard-wall confining potential shape of the four terminal junction with
round corners. The roundness of each corner can be controlled independently by varying the radii of
the circular corners, $R_{\mathrm{C1}}$ and $R_{\mathrm{C2}}$.
In a region centered on the scattering area we have a SO coupling which is described by the
symmetrized Rashba Hamiltonian
\begin{equation}
 H_{\mathrm{R}}(x,y)=\frac{1}{2\hbar}\left\lbrace\alpha(x,y),(p_y\sigma_x-p_x\sigma_y)\right\rbrace.
 \label{eq:RashbaHamilton}
\end{equation}
Here $\alpha(x,y)$ is the Rashba SO coupling strength which is turned on smoothly at the edges of the
scattering region. Having Rashba SO coupling only in a finite area of the sample can be achieved by
using metallic gate on top of the 2DEG \cite{Engels:prb1997}. The Hamiltonian is discretized on a 
square lattice using the finite difference method, resulting in an infinitely large matrix. The
corresponding matrix equation for the retarded Green's function matrix of the system is
\begin{equation}
(E\mathbf{I}-\mathbf{H})\mathbf{G}^{\mathrm{r}}=\mathbf{I}.
\label{eq:InfiniteGreen}
\end{equation}
The scattering region part of the Hamiltonian Eq.\ (\ref{eq:InfiniteGreen}) can be
made finite by treating the leads as self-energies in the standard way \cite{Datta:book1995,Ferry:book1997}. The leads have no SO coupling. The finite version of the matrix
equation is
\begin{equation}
 \big(E\mathbf{I}-\mathbf{H_{\mathrm{s}}}-\sum_j\mathbf{\Sigma}_j\big)\mathbf{G}^{\mathrm{r}}_{\mathrm{s}}
 =\mathbf{I}.
 \label{eq:FiniteGreen}
\end{equation}
Here the self-energy of lead $j$ is $\mathbf{\Sigma}_j$, $\mathbf{H_{\mathrm{s}}}$ is the Hamiltonian of the
scattering region and $\mathbf{G}^{\mathrm{r}}_{\mathrm{s}}$ is the retarded Green's function.
All the matrices are $2N_xN_y \times 2N_xN_y$ matrices. In order to save computational power only the necessary
Green's function matrix elements are calculated with the recursive Green's function method
\cite{Ferry:book1997}. From the Green's functions both the spin conductance and spin-densities can
be calculated. The $z$ spin density for an bias of $eV_0$ is
\begin{equation}
 \langle \mathbf{S}_z(\mathbf{r}) \rangle=\frac{\hbar}{2}\int_{E_{\mathrm{F}}-eV_0/2}^{E_{\mathrm{F}}+eV_0/2}
 \frac{dE}{2\pi i}\mathrm{Tr}_{\mathrm{spin}}
 \left[\mathbf{\sigma}_z \mathbf{G}^<(\mathbf{r},\mathbf{r};E)\right],
 \label{eq:SpinDensity}
\end{equation}
where $\sigma_z$ is the Pauli matrix for the $z$ direction, 
\begin{equation}
 \mathbf{G^{<}}(\mathbf{r},\mathbf{r};E)
 =i\left[\mathbf{G}^{\mathrm{r}}_{\mathrm{s}}(E)
 \left(\sum_j f_j(E)\mathbf{\Gamma}_j(E)\right)
 \mathbf{G}^{\mathrm{a}}_{\mathrm{s}}(E)\right]_{\mathbf{r},\mathbf{r}}
 \label{eq:LesserGreen}
\end{equation}
is the lesser Green's function\cite{Haug:book2008}, $\mathbf{\Gamma}_j=-2\mathrm{Im}(\mathbf{\Sigma}_j)$, and
$f_j(E)=1/[\exp(\beta(E-\mu_j))+1]$ is the Fermi function in lead $j$ which has the chemical potential
$\mu_j=E_{\mathrm{F}}+eV_j$. We define the spin-dependent conductance between the leads $q$ and $p$ as
\begin{equation}
 G^{\sigma\sigma'}_{pq}=\frac{e^2}{h}\left[\mathrm{Tr}
 \left[\mathbf{\sigma_z}\mathbf{\Gamma}_q\mathbf{G}^{\mathrm{r}}_{\mathrm{s}}
 \mathbf{\Gamma}_p\mathbf{G}^{\mathrm{a}}_{\mathrm{s}}\right]
 \right]_{\sigma,\sigma'}
 \label{eq:SpinConductance}
\end{equation}
Using the Landauer B\"{u}ttiker formula we write the linear response of the $\sigma=\uparrow,\downarrow$ spin
depended current through lead $p$ as
\begin{equation}
 I_p^{\sigma}=\sum_{q,\sigma'} G_{qp}^{\sigma,\sigma'}(V_p-V_q).
 \label{eq:SpinCurrent}
\end{equation}
%

\section{Numerical simulations}

The dimension of the scattering area is set within a $L_x\times L_y$ box shaped area, where $L_x=L_y=600$ nm.
This area is discretized on a $102\times102$ point grid with constant mesh size 
$a = 5.88$ nm and the effective electron mass is set as $m^*=0.0447m_{\mathrm{e}}$, which is a reasonable value for
 Ga$_{1-x}$In$_x$As alloys\cite{Levinshtein:book1996}. In the simulations all lengths where scaled in
the mesh size $a$ and energies in the tight-binding hopping term $t=\hbar^2/2m^*a^2 = 24.6$
meV which results from the finite difference discretization of the Hamiltonian in Eq.\ 
(\ref{eq:MainHamilton}). Four leads connect to the system, two main leads with width
$W_y=41a = 241$ nm (leads 1 and 2) and two transverse leads with width $W_x=18a = 106$ nm
(leads 3 and 4). In the following, the energy will be scaled in the lowest transverse energy
of the main leads, $E_0 = 0.145$ meV$ = 5.89\cdot10^{-3}$t. Similarly we also define
$E_{\mathrm{t}} = 5.46 E_0$ which is the lowest transverse energy of the transverse leads.
The effect of Rashba SO coupling is turned on smoothly over a $10 a =  58.8$ nm
strip at the edge of the scattering area, see the hatched area in Fig.\ \ref{fig:scematic}, to its 
full strength $\alpha_0$. Between the two main leads we
apply a bias $eV_0=1E_0$. This means that the chemical potentials at the leads measured from $E_{\mathrm{F}}$ will
be $eV_1=eV_0/2$, $eV_2=-eV_0/2$ and $eV_3=eV_4=0$. According to Eq.\ (\ref{eq:SpinCurrent}) the spin
depended current ($\sigma=\uparrow,\downarrow$) is then 
\begin{equation}
 I_{t}^{\sigma}=\left(G_{t2}^{\sigma\sigma}+G_{t2}^{\sigma\overline{\sigma}}
  		     -G_{t1}^{\sigma\sigma}-G_{t1}^{\sigma\overline{\sigma}}\right)
 \frac{V_0}{2}\qquad t=3,4.
 \label{eq:TransverseSpinCurrent}
\end{equation}
We define the spin Hall conductance in transverse lead $t$ as
\begin{equation}
 G_{\mathrm{sH}_t}=\frac{\hbar}{2e}\frac{I_t^{\uparrow}-I_t^{\downarrow}}{V_0} \qquad t=3,4.
\end{equation}
Note that in our scenario the total charge current through the transverse leads is always zero because of how
we define the bias over the leads. Throughout the paper the full strength of the Rasbha SO coupling will be 
set as $\alpha_0= 10$ meV nm$= 2at_{\mathrm{SO}}$, where $t_{\mathrm{SO}} = 3.46\cdot10^{-2} t$.  This value of Rashba SO coupling is typical for Ga$_{1-x}$In$_x$As alloys and is also in
the regime where we begin to see strong and interesting features in spin conductance through the transverse lead.

\subsection{Ballistic system}
\begin{figure}[htp]
 \begin{center}
 \includegraphics[width=0.49\textwidth]{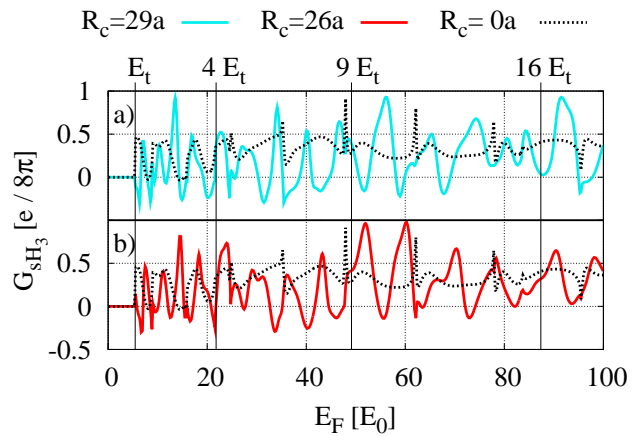}
 \end{center}
 \caption{
 (Color online) Spin conductance through lead 3 plotted as a
 function of Fermi energy. Figure a) correspond to the four terminal
 shape where $R_{\mathrm{C}}=29 a$, b) correspond to
 $R_{\mathrm{C}}=26 a$. In both figures the spin conductance through
 a $R_{\mathrm{C}}=0 a$ system (sharp corners) is plotted for comparison.
 The bottom of the energy bands of transverse leads
 (with energy scaled in $E_{\mathrm{t}}$) have been marked onto the figure.
 Note that the $y$ scale is unique for each of the figures.
 }
 \label{fig:threeCurvRadius}
\end{figure}

We will start by showing the spin Hall conductance through the transverse leads in
a clean system. In Fig.\ \ref{fig:threeCurvRadius} the spin Hall conductance
through lead 3,\ $G_{\mathrm{sH}_3}$, is plotted as a function of Fermi
energy $E_{\mathrm{F}}$ for three types of shapes. Fig.\ \ref{fig:threeCurvRadius} a) corresponds to 
shape with $R_{\mathrm{C}}=29 a$, Fig.\ \ref{fig:threeCurvRadius} b) to $R_{\mathrm{C}}=26 a$.
In both figures the the spin conductance for a sharp corner system ($R_{\mathrm{C}}=0 a$)
is plotted for comparison.
Here we use $R_{\mathrm{C}}=R_{\mathrm{C}1}=R_{\mathrm{C}2}$ when all the corners have the same curvature. 

In Fig.\ \ref{fig:threeCurvRadius} a) and b) we see that the spin conductance curves corresponding 
to the round corner system shows a series of minima and maxima. These extrema are most likely related to the 
resonance states in the scattering area which we will discuss below.

\begin{figure}[htp]
\begin{center}
 \centering
 \includegraphics[width=0.45\textwidth]{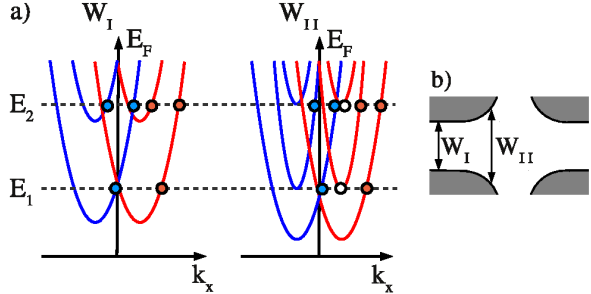}
\end{center}
 \caption{
 (Color online) Schematic picture showing the effective shift in the
 transverse Rasbha energy bands, see a), when the system gets wider,
 see b). In a) the left energy bands corresponds to the system where
 the width is $W_{\mathrm{I}}$ and the right energy bands correspond
 to the system where the width is $W_{\mathrm{II}}$. Here $k_x$ is the
 longitudinal k-vector of the electrons. The red lines correspond to spin up states
 and the blue lines to spin down states. At Fermi energies $E_1$ and $E_2$
 we get additional states with low longitudinal velocities ($v_x\approx 0$),
 marked with white dots, in addition to the states with
 positive velocities, marked with red and blue dots. In the terminal area
 the electrons from the original positive longitudinal velocity states
 can scatter into these newly open states. 
 }
 \label{fig:ScematicBandShift}
\end{figure}

The resonance states form when
the leads get adiabatically larger which results in a downward shift in the transverse energy
bands of the leads\cite{Baranger:prb1991}, see Fig.\ \ref{fig:ScematicBandShift}. This causes new
states to become available at the Fermi energy which have very low longitudinal velocity ($v_x\approx 0$).
Near the transverse leads the adiabatic approximation ceases to apply and the electron scatter
in to different states, including the newly opened low longitudinal velocity states which give rise
to conductance extrema in the transverse leads.

The sharp corner $G_{\mathrm{sH_3}}$ curve, shown both in Fig.\ \ref{fig:threeCurvRadius} a) and b),
behaves differently. At high enough energies ($E_F > 30E_0$) the curve changes slowly with energy and
has no significant resonance peaks, apart from isolated peaks due to divergence in the density of
states at energies corresponding to the band bottom in the longitudinal leads.  The observed behaviour
of the sharp corner $G_{\mathrm{sH_3}}$ is due to the strong scattering by the sharp corners.
An electron coming in from the left, with a definite $k$-value, gets scattered into all possible states,
with the same energy,
when it enters the scattering region. For higher energies these states are many, all corresponding
to different effective magnetic fields which tend to average out the spin-signal. Note that for low
energies we still see peaks and minima since only a few $k$-states are available at such low energies.

\begin{figure}[tbhp!]
 \includegraphics[width=0.49\textwidth]{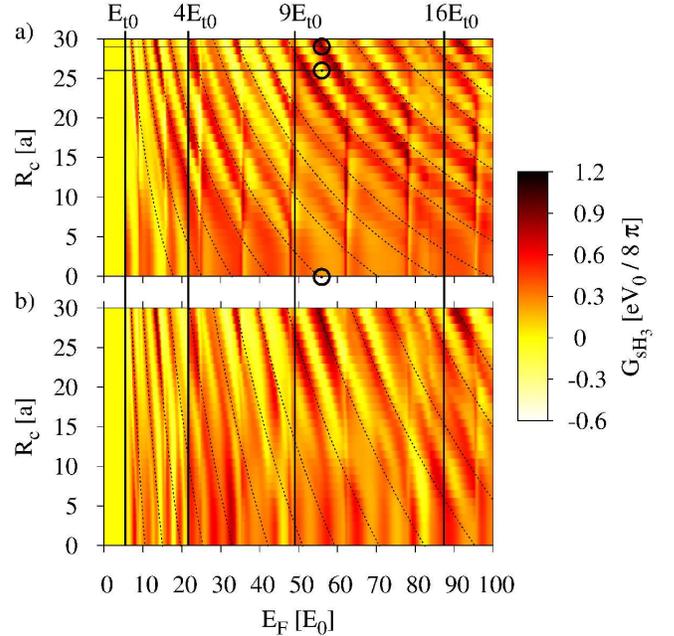}
 \caption{
 (Color online) Spin conductance through lead 3 plotted as a function of radii
 of the corners and Fermi energy. In a) all corners are equally varied while
 in b) two corners are varied equally, the $R_{\mathrm{C}1}$ corners, and two
 corners are kept fixed, $R_{\mathrm{C}2}=30 a$. The black dashed lines show
 the shift of $G_{\mathrm{sH}}$ maxima in energy according to Eq.\ (\ref{eq:EnergyDrift}).
 The horizontal black lines mark where the spin conductance through lead 3 in
 Fig.\ \ref{fig:threeCurvRadius} a) and b) lie and the black circles show the
 ($R_{\mathrm{C}}$, $E_{\mathrm{F}}$) coordinates of the spin densities shown in Fig.\
 \ref{fig:spinDensity}. The vertical black lines mark the bottom of the
 transverse energy bands scaled in $E_t= 5.46 E_0$.}
 \label{fig:spinCurrVsCorners}
\end{figure}

Comparing the round corner curves in Fig.\ \ref{fig:threeCurvRadius} a) and b) we see that the extrema
at $E_F=56E_0$ in
Fig.\ \ref{fig:threeCurvRadius} a) has shifted to $E_F=60E_0$ in Fig.\ \ref{fig:threeCurvRadius} b). 
Other extrema in $G_{\mathrm{sH_3}}$ seem also to have been shifted by the change in corner
radius $R_{\mathrm{C}}$.
To explore in more detail this shift of extrema in $G_{\mathrm{sH_3}}$ with change in curvature we make a
surface plot of the spin Hall conductance through lead 3, $G_{\mathrm{sH}_3}$, as a function of Fermi
energy varied from $0 E_0$ to $100 E_0$ and $R_{\mathrm{C}}$ varied from $0 a$ (sharp corners) to
$30 a$ with interval of $1 a$. The result can be seen in Fig.\ \ref{fig:spinCurrVsCorners} a).
There we see that for low corner curvatures ($R_{\mathrm{C}}\lesssim10 a$) the spin conductance 
becomes smeared out as was discussed above.

For higher corner curvatures ($R_{\mathrm{C}}\gtrsim10 a$) we get a spin conductance curve which shows
a series of minima and maxima. Extrema in $G_{\mathrm{sH}_3}$ also begin to shift in energy with increasing
corner curvature. For smoother corners the transport gets more adiabatic and the extrema in $G_{\mathrm{sH}_3}$
will be more influenced by geometric resonances, i.e.\ changes in shape will be adiabatically translated into
a shift in the energy bands, that will affect the transport. Changing the curvature of the scattering
region effectively changes its area. For smooth enough corners,the shape of the system is not affected
by changing the curvature, only its size, which results in a universal shift of all bands in the scattering
area. This does not apply to the sharp corners, where a small change in curvature can results in great
change in scattering properties.

The effective diameter of the scattering region is $D=(\sqrt{2}-1)(R_{\mathrm{C}1}+R_{\mathrm{C}2})+\sqrt{W_x^2+W_y^2}$,
see Fig.\ \ref{fig:scematic}.
The energy shift of a given $G_{\mathrm{sH}_3}$ extrema will be $E_\mathrm{C}\propto D^{-2}$. 
The prefactor is not known, but this does not matter as we will show here below.
We can pick a reference energy $E_{\mathrm{C}_0}$, with reference size 
$D_0=(\sqrt{2}-1)(R_{\mathrm{C}1_0}+R_{\mathrm{C}2_0})+\sqrt{W_\mathrm{A}^2+W_\mathrm{B}^2}$, 
on the same extrema in $G_{\mathrm{sH_3}}$. By dividing $E_{\mathrm{C}}$ with this reference
energy $E_{\mathrm{C}_0}$ we get rid of the unknown prefactor in $E_{\mathrm{C}}$ and obtain
\begin{equation}
 E_\mathrm{C}=\left[\frac{(\sqrt{2}-1)(R_{\mathrm{C}1_0}+R_{\mathrm{C}2_0})+\sqrt{W_x^2+W_y^2}}
                   {(\sqrt{2}-1)(R_{\mathrm{C}1}+R_{\mathrm{C}2})+\sqrt{W_x^2+W_y^2}}\right]^2
                   E_{\mathrm{C}_0},
 \label{eq:EnergyDrift}
\end{equation}
which describes the shift of the energy levels in the scattering area as a function of the shape of
the scattering area.

In Fig.\ \ref{fig:spinCurrVsCorners} a) we have plotted curves, shown with black dashed lines, with
reference points at $R_{\mathrm{C}_0}=R_{\mathrm{C}1_0}=R_{\mathrm{C}2_0}=30a$ on all the 
spin conductance maxima.  As can be seen for $R_C\gtrsim 10 a$ the shift of the maxima curves fits well
to the shift that we expect for adiabatic change in size of the scattering area. For 
$R_{\mathrm{C}}\lesssim 10 a$ scattering from the corners gets more dominating, which is to be 
expected since the system is out the adiabatic regime and the sensitivity to the corners takes over.

A shape where we keep $R_{\mathrm{C}2}=30a$ fixed and only vary $R_{\mathrm{C}1}$ was also examined, see 
Fig.\ \ref{fig:spinCurrVsCorners} b). Here we also plot curves based on Eq.\ (\ref{eq:EnergyDrift}),
shown with black dashed lines, with reference points at $R_{\mathrm{C}1}=30a$ on all the maxima curves.
We notice that the for low $R_{\mathrm{C}1}$ corner curvatures, 
$R_{\mathrm{C}1}\lesssim 10 a$, we get somewhat more structure than for low $R_{\mathrm{C}}$
in the equal corner shape. This is due to that only two corners are contributing to the
sharp corner scattering. We also notice that for $R_{\mathrm{C}1}\gtrsim 10 a$ the shift in spin
conductance maxima follows the expected shift calcuated from Eq.\ (\ref{eq:EnergyDrift}).

\begin{figure}[tbhp!]
 \centering
 \includegraphics[width=0.40\textwidth]{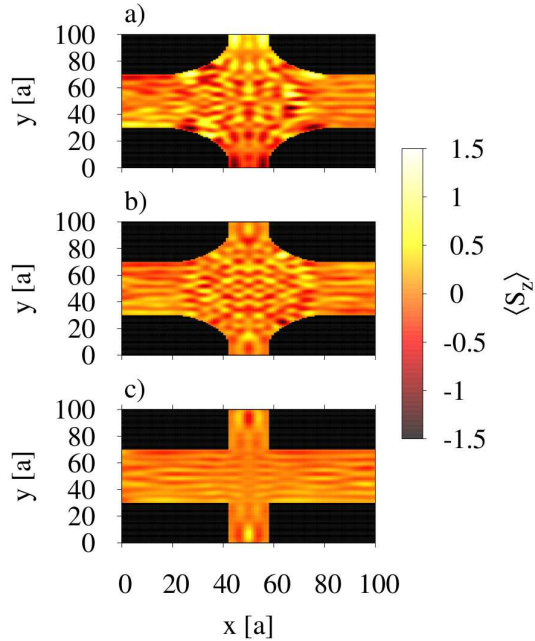}
 \caption{
 (Color online) Density of the spin $z$ component in the ballistic system
 at $E_{\mathrm{F}}=56E_0$ for three different geometries. The shape in a) has
 $R_\mathrm{C}=29 a$, the shape in b) has $R_\mathrm{C}=26 a$,
 and the shape in c) has $R_\mathrm{C}=0 a$ (sharp corners). These
 ($E_{\mathrm{F}},R_{\mathrm{C}}$) values correspond to the spin Hall values marked
 with black circles in Fig.\ \ref{fig:spinCurrVsCorners} a).}
\label{fig:spinDensity}
\end{figure}

To examine the effect of corner curvature better we also plot the density of the $z$ spin component
in the scattering region for three types of all equal corner shapes $R_{\mathrm{C}}=29 a$, see Fig.\ 
\ref{fig:spinDensity} a), $R_{\mathrm{C}}=29 a$, see Fig.\ \ref{fig:spinDensity} b), and
 $R_{\mathrm{C}}=0 a$, see Fig.\ \ref{fig:spinDensity} c). For all three
types we set the bias as $eV_0=1E_0$ and temperature at $T=1K$. 

In Fig.\ \ref{fig:spinDensity} a) we have stronger spin densities along the corner edges and within 
the transverse leads than in Fig.\ \ref{fig:spinDensity} b). Also the spin densities averaged over the transverse
leads in Fig.\ \ref{fig:spinDensity} b) is lower compared to equivalent averaging in
Fig.\ \ref{fig:spinDensity} a). These densities
correspond to spin conductance strength marked with black circles in 
Fig.\ \ref{fig:spinCurrVsCorners} a) and as expected the more negative spin conductance yields more 
negative spin density averaged over the transverse leads. For comparison we also include spin density in the sharp
corner system, see Fig.\ \ref{fig:spinDensity} c). We see that the spin density in the sharp corner system is more
smeared out than in the round corner systems, as could be expected in light of
previous discussion about the sharp corner system. Note though that these spin densities seen
in Fig.\ \ref{fig:spinDensity} are not large, only up to $\approx1.3\cdot10^{-6}$ $\hbar/2$ per nm$^2$.

The results presented in this section suggest that polarization of the spin current through the transverse
leads can be effectively controlled by tuning the curvature of the corners in the scattering area. This could
be realised by using e.g. finger gates\cite{Liang:prb2004}.
\subsection{Quasi-ballistic system}
To test how robust the $G_{\mathrm{sH}}$ extrema are, we add impurity effects to our model.
Two different methods of including impurity effects were studied. The first
method which we will consider involves adding a few randomly distributed Gaussian shaped impurities to
the sample. The second method which we considered is the
Anderson impurity method which has already been extensively used for similar simulations
\cite{Anantrac:jrb1998,Nikolic:prb2005:b,Nikolic:prl2005,Garelli:prb2009,Anantrac:jrb1998,
Duckheim:prb2009,Grincwajg:jpcm1997}.

\subsubsection{Static impurities}
\begin{figure}[tbhp!]
 \includegraphics[width=0.49\textwidth]{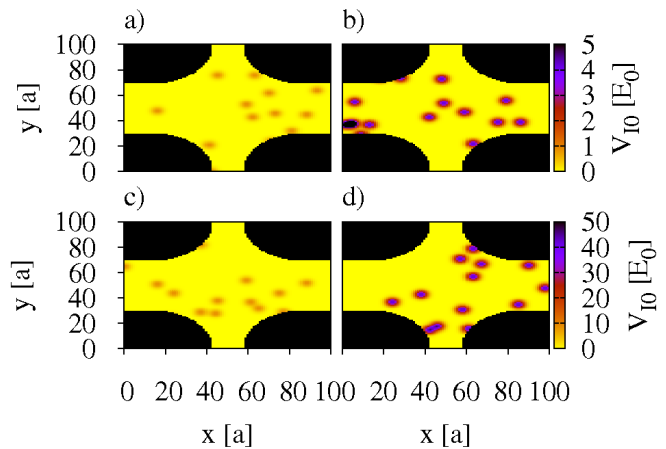}
 \caption{
 (Color online) Scattering shape using $R_{\mathrm{C}}=29a$ corner radii 
 and with $N_{\mathrm{I}}=20$ static impurities added on random locations in the whole box of the scattering area.
 All impurities are set to  have range $\Gamma_{\mathrm{I}}=2a$, and the impurity strength is set
 at $V_{\mathrm{I}_0}=1E_0$ for impurities in a), $V_{\mathrm{I}_0}=5E_0$ for impurities in 
 b), $V_{\mathrm{I}_0}=10E_0$ for impurities in c), and $V_{\mathrm{I}_0}=50E_0$ for impurities in d).
 }
 \label{fig:ImpurityPot}
\end{figure}

\begin{figure}[tbhp!]
 \includegraphics[width=0.45\textwidth]{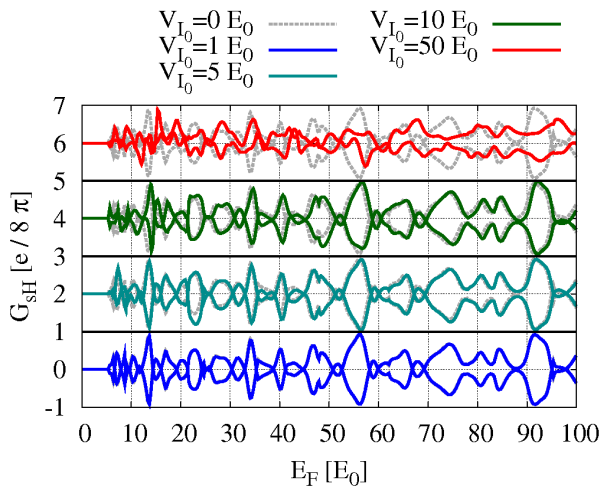}
 \caption{
 (Color online) The spin conductance through lead 3 and 4 as a function of
 energy for the cases shown in Fig.\ \ref{fig:ImpurityPot} compared
 to the spin conductance of the clean system, see Fig.\
 \ref{fig:threeCurvRadius} a). Note that the $G_{\mathrm{sH}}$ scale is
 shifted incrementally for each $V_{\mathrm{I}_0}$ curve from the
 $V_{\mathrm{I}_0}=1E_0$ curve.
 }
\label{fig:ImpurityCond}
\end{figure}

Phenomena such as crystal defects can introduce extra potential bumps in the otherwise 
uniform potential background.
These bumps can be described as static Gaussian shaped impurities in the scattering region which we add randomly to our
scattering region via the term
\begin{equation}
 V_{\mathrm{I}}(x,y)=V_{\mathrm{I}_0}\sum_{i=1}^{N_{\mathrm{I}}}
    \exp\left(-\frac{(x_i-x)^2+(y_i-y)^2}{2\Gamma_{\mathrm{I}}^2}\right),
 \label{eq:ImpurityPotential}
\end{equation}
which is added to the Hamiltonian in Eq.\ (\ref{eq:MainHamilton}).
Here $V_{\mathrm{I}_0}$ is the potential height of the potential bump, $N_{\mathrm{I}}$ is the number 
of impurities, $(x_i,y_i)$ the center point of each impurity and $\Gamma_{\mathrm{I}}$ impurity range.
We would expect the number of these impurities to be just a few percent of the number of donors. 
We choose 2\% which corresponds to roughly  $N_{\mathrm{I}}=20$ Gaussian impurities in an $600\times600$ nm$^2$
area and a delta donor density of $\approx 3\cdot10^{11}$ cm$^{-2}$.
We are interested in seeing how strong these Gaussian shaped impurities must be to have significant
effect on the spin hall conductance. 
To accomplish this four cases of impurity strength were studied:
$V_{\mathrm{I}_0}=1E_0$, $V_{\mathrm{I}_0}=5E_0$, $V_{\mathrm{I}_0}=10E_0$, and  $V_{\mathrm{I}_0}=50E_0$,
see Fig.\ \ref{fig:ImpurityPot}. These values of impurity strength are roughly the same order
of magnitude as a screened and unscreened point charge 17 nm away from a 2DEG in Ga[Al]As
heterostructure\cite{Ihn:book2010}.
The effect these impurities have on the spin Hall conductance in leads
3 and 4 can be seen in Fig.\ \ref{fig:ImpurityCond}. There we see that up to impurity strength $V_{\mathrm{I}_0}=10E_0$ the spin conductance is rather
stable and that spin conductance through lead 3 and 4 is symmetric around zero.

\subsubsection{Anderson type impurities}
\begin{figure}[tbhp!]
 \includegraphics[width=0.49\textwidth]{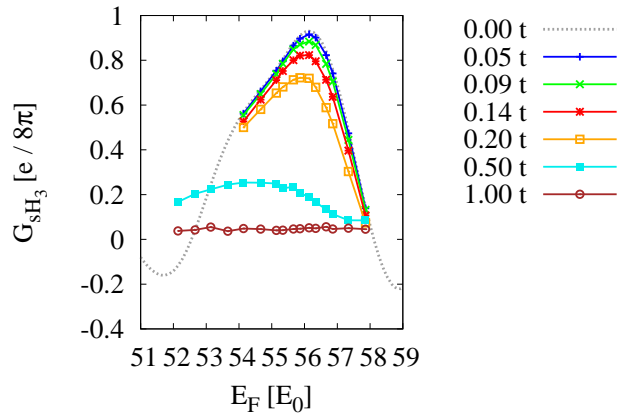}
 \caption{
 (Color online) The spin conductance through lead 3 as a function of
 Fermi energy for a range of disorder strengths $U$. To compare we also
 plot he spin conductance of the ballistic system which we mark with $U=0t$.
 These results have been averaged over 1000 runs and the disorder strength
 is scaled in the tight binding hopping term $t = 171 E_0$.}
 \label{fig:AndersonImpurityCond}
\end{figure}

The Anderson impurity potential adds to each point in the scattering area a random value of potential
energy in the range $[-U/2,U/2]$, where $U$ describes the disorder strength.
To study the effect of the Anderson type impurities on $G_{\mathrm{sH}}$ we focus on a extrema
in $G_{\mathrm{sH_3}}$ at $E_\mathrm{F} = 56 E_0$ for a $R_{\mathrm{C}}=29a$ scattering shape,
see Fig.\ \ref{fig:threeCurvRadius} a) . In Fig.\ \ref{fig:AndersonImpurityCond}
the spin Hall conductance is plotted for impurity strength ranging from $U=0.05 t$ to $U=1.0 t$. Each data point is
averaged over 1000 impurity configurations. For comparison we include the result for the ballistic system,
seen also in Fig.\ \ref{fig:threeCurvRadius} a). In Fig.\ \ref{fig:AndersonImpurityCond} we see that the
structure of the spin conductance is stable up to $U=0.2 t=34.2E_0$ although its amplitude diminishes
and the structure shifts with increasing disorder strength. For $U=0.5t=85.5E_0$ the spin conductance
seems to be disappearing as has been shown for such large disorder values\cite{Nikolic:prb2005:c}.
%

\section{Conclusions and discussion}
In this paper the effect of scattering area shape on spin conductance in a four-terminal 2DEG
structure with Rashba SO coupling is studied numerically.
The change in the scattering region was implemented by modifying the curvature of the corners between
different leads.
Both the ballistic regime and the quasi-ballistic regime were considered. It is found that above a certain
curvature the spin conductance,  $G_{\textrm{sH}}$, shows a series of minima and maxima,
as a function of energy. These minima and maxima are shifted in energy with increasing curvature.
We propose a relation between this shift of the extrema and the
change in the size of the scattering area when the curvature is increased.
This is due to the increased energy for smaller
scattering areas and describes the observed shift of the extrema reasonably well.
With this relation the polarization of the spin current through the transverse
leads could effectively be controlled by tuning the curvature of the corners in the scattering area.

In addition, we have also test the robustness of the spin conductance extrema by adding impurities to the
scattering area. Static and Anderson type impurities were considered. The amplitude of
the spin conductance extrema are suppressed with increasing impurity strength but the structure is shown to
be stable under reasonable values of impurity strength, i.e. when the system is quasi-ballistic.

\acknowledgments{We gratefully acknowledge helpful discussions with Mathias D\"{u}ckheim and Andrei Manolescu.
This work was supported by the Icelandic Science and Technology Research
Program for Postgenomic Biomedicine, Nanoscience and Nanotechnology, the Icelandic Research Fund, and the Research
Fund of the University of Iceland.}


\end{document}